\documentclass[a4paper,fleqn,usenatbib,useAMS]{mnras}

\usepackage[T1]{fontenc}
\usepackage{ae,aecompl}

\usepackage{graphicx}	% Including figure files
\usepackage{amsmath}	% Advanced maths commands
\usepackage{amssymb}	% Extra maths symbols

\usepackage{times}

\title[Magellanic Miras with \textit{Gaia} DR1]{The Clouds are breaking: tracing the Magellanic system with \textit{Gaia} DR1 Mira variables}

\author[A. J. Deason et al.]{
Alis J. Deason$^{1}$\thanks{E-mail: alis.j.deason@durham.ac.uk},
Vasily Belokurov$^{2}$, Denis Erkal$^{2}$, Sergey E. Koposov$^{2}$,
\newauthor Dougal Mackey$^{3}$\\
$^{1}$Institute for Computational Cosmology, Department of Physics, University of Durham, South Road, Durham DH1 3LE, UK\\
$^{2}$Institute of Astronomy, University of Cambridge, Madingley Road, Cambridge CB3 0HA, UK\\
$^{3}$Research School of Astronomy and
  Astrophysics, Australian National University, Canberra, ACT 2611,
  Australia
}

\date{Accepted XXX. Received YYY; in original form ZZZ}

\pubyear{2017}

\begin{document}
\label{firstpage}
\pagerange{\pageref{firstpage}--\pageref{lastpage}}
\maketitle

\begin{abstract}
We exploit the first data release from the \textit{Gaia} mission to identify candidate Mira variables in the outskirts of the Magellanic Clouds. The repeated observations of sources during the initial phase of the \textit{Gaia} mission is used to identify stars that show signs of variability. This variability information, combined with infrared photometry from 2MASS and WISE, allows us to select a clean sample of giants in the periphery of the LMC. We find evidence for Miras surrounding the LMC out to $\sim20$ deg in all directions, apart from the North-West quadrant. Our sample does not generally follow the gas distribution of the Magellanic system; Miras are notably absent in the gaseous bridge between the LMC and SMC, but they are likely related to the \textit{stellar} RR Lyrae bridge reported by \cite{rrl}. The stellar stream discovered by \cite{mackey16} to the North of the LMC is almost perfectly delineated by our Mira variables, and likely extends further East toward the Galactic plane. The presence of an intermediate-age population in this stream advocates an LMC disc origin. We also find a significant excess of Miras to the East of the LMC; these more diffusely distributed stars are likely stripped SMC stars due to interactions with the LMC.  Miras are also identified in regions of the sky away from the Clouds; we locate stars likely associated with known massive substructures, and also find potential associations with stripped SMC debris above the Galactic plane.
\end{abstract}


\section{Introduction}

The Large and Small Magellanic Clouds (LMC and SMC) are a pair of massive, star-forming dwarf galaxies located $50-60$ kpc from the Galactic centre. A vast network of gaseous material surrounds the Clouds, comprising the Magellanic Stream, Bridge and Leading Arm (see e.g \citealt{putman03, bruns05, nidever08}). The structure extends over an impressive 200 deg on the sky, and illustrates the complex, interacting past of the dwarf pair with each other, and the Milky Way (MW). The proximity of the Clouds allows us to scrutinize both their gaseous \textit{and} stellar components. Thus, this pair of satellite dwarfs provide the closest, and most detailed example of a three-body train wreck in the local universe (see \citealt{donghia16} for a recent review). 

The dynamical history of the LMC and SMC, and the subsequent formation of the gaseous Magellanic complex, has been modeled extensively in the literature (e.g, \citealt{gardiner96, mastropietro05, connors06, besla07, diaz11, guglielmo14}). Historically, two competing scenarios --- tidal interactions and ram-pressure stripping --- were brought forward to explain the formation of the Magellanic stream \citep{fujimoto76, lin77, meurer85, moore94}. However, as the observational data surrounding the Clouds has grown in volume and intricacy, so have the numerical models. Current state-of-the-art models that take into account the latest orbital constraints on the cloud dynamics \citep{kalli13}, show that the gaseous Magellanic structures likely formed from interactions between the LMC and SMC, and are additionally shaped by the MW tides and hot halo (see e.g. \citealt{besla12, besla13, diaz11,diaz12, hammer15}).

The outskirts of the LMC and SMC are invaluable testing grounds for the various formation mechanisms of the Magellanic system. In particular, the long-awaited confirmation of the stellar counterpart to the Magellanic stream, which is predicted by all tidal models, will likely be uncovered in the low surface brightness regions surrounding the Clouds (e.g \citealt{weinberg00}). Furthermore, evidence for past interactions between the two dwarfs, and disturbances due to the MW tidal field should be more apparent in these low-density regions. For example, \cite{mackey16} and \cite{besla16} recently showed that stream-like structures found in the outer LMC disc can be caused by MW tides and/or repeated interactions with the SMC. 

The nature and extent of the LMC's diffuse stellar halo component is still under debate. \cite{majewski09} argue that the presence of remote LMC stars $\sim 20$ deg from the LMC centre are indicative of a pressure supported halo. However,  deep photometric surveys by \cite{saha10}, \cite{balbinot15} and \cite{mackey16} find star counts consistent with a very extended LMC disc, which stretches out to at least $\sim15-16$ deg from the LMC centre. Hence, if the halo component does exist, it may not dominate the star counts until several tens of degrees from the LMC. \cite{belokurov16} probed further out from the LMC using old, blue horizontal branch stars and uncovered a enormous, lumpy LMC stellar halo reaching out to at least $\sim 30$ deg. These results indicate that the LMC is much more massive than previously recognized, in agreement with the recent measurement by \cite{jorge_lmc}, which has important implications for models of the Clouds dynamical history. Finally, it is worth noting that models of LMC-SMC interactions predict that stripped stellar SMC material will be diffusely distributed about the disc of the LMC \citep{diaz12, besla13}, so some LMC halo stars could have an SMC origin. In fact, evidence for SMC populations in the LMC have already been observed. \cite{olsen11} found a kinematically distinct population of low metallicity ($[\mathrm{Fe/H}] \sim -1.25 $) giant stars in the LMC, and the authors argue that these stars likely originated in the SMC.
 
It is clear that the relatively unexplored outer regions of the LMC and SMC likely contain a gold mine of information pertaining to the origin and formation of the Magellanic system. However, observations of the low-density stellar components of the Clouds remains a challenge. The relatively low Galactic latitude ($b \sim -30$ deg) of the LMC means that its outskirts are dominated by regions of high foreground stellar density. Thus, at distances of $\sim 50-60$ kpc, intrinsically bright stellar tracers that can be \textit{cleanly} selected in regions of high extinction are required. 

Asymptotic Giant Branch (AGB) stars have often been used in the literature to the study the LMC and SMC (e.g, \citealt{blanco80, kunkel97, vandermarel01, cioni06}). These red stars can be efficiently selected close to the MW disc plane using a combination of near- and far-infrared photometry \citep{koposov15}. AGBs also have very bright absolute magnitudes, making them ideal targets for follow-up spectroscopy. However, these tracers are not abundant, and, even after photometric selection, the foreground stellar contamination can become overwhelming in regions of low LMC/SMC density.

Most giant stars are variables; their intrinsic brightness periodically changes over time due to thermal pulsations. The variability of giants can be used to distinguish them from foreground dwarf stars, which otherwise have similar photometric properties. In this work, we exploit their variable nature to identify AGBs in the outskirts of the LMC and SMC. We develop a novel technique using the first data release (DR1) of the \textit{Gaia} mission \citep{gaia16} to select variable giants (i.e. Mira variables). \textit{Gaia} is an astrometric mission that will measure positions, parallaxes and proper motions for hundreds of millions of stars in our Galaxy. We combine the first data release of this unprecedented mission, with all-sky infrared surveys (2MASS and WISE) to cleanly select Miras in the vicinity of the LMC.

The paper is arranged as follows. In Section \ref{sec:sel}, we describe our selection of Mira variables in the outskirts of the LMC.  Here, we use infrared photometry from 2MASS and WISE combined with variability information from \textit{Gaia} DR1 to select our sample. The stellar features that we uncover are described in Section \ref{sec:periph}, and we discuss using Miras to trace other MW halo substructures in Section \ref{sec:subs}. Finally, we summarise our main conclusions in Section \ref{sec:conc}.

\begin{figure*}
  \begin{center}
    \includegraphics[width=15cm, height=10cm]{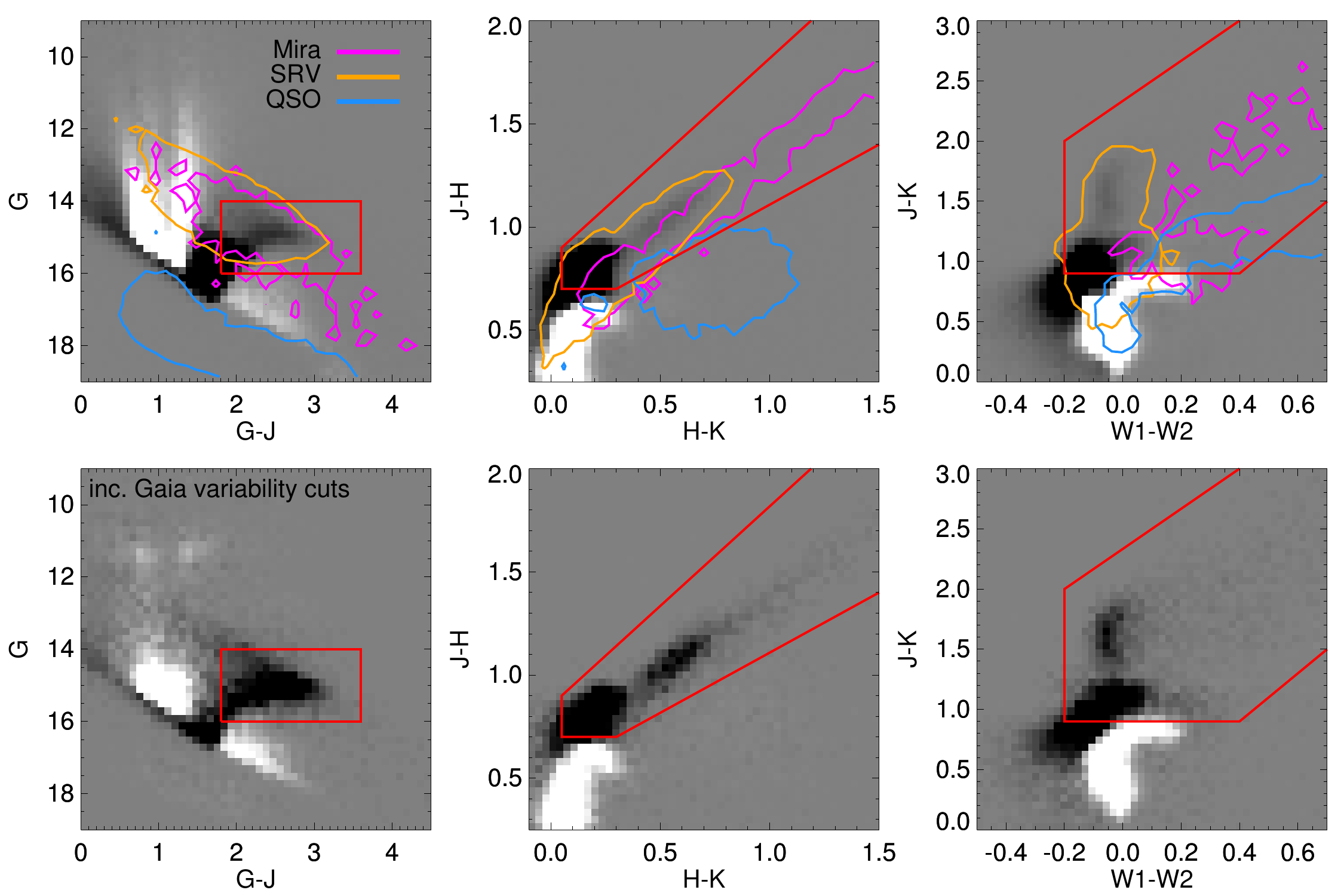}
    \caption{\label{fig:cmd} \small Colour-magnitude diagram (CMD) and colour-colour plots used to select Mira variables in the vicinity of the LMC. We show the difference between stars selected within 12 degrees of the LMC and the foreground stellar distribution. Here, the ``foreground'' are objects between 25 and 50 degrees from the LMC and outside of 10 degrees of the SMC. The foreground is normalised, and subtracted from the LMC field. Thus, the gray shading shows equality with the foreground and the black (white) indicates where the LMC (foreground) stars are dominant.  The left panel shows the G-J, G CMD, where the G band is from \textit{Gaia} and the J band is from 2MASS. In the middle and right panels we show the 2MASS H-K, J-H and 2MASS+WISE W1-W2, J-K colour-colour diagrams, respectively. Note that we only show objects with WISE magnitudes W1 < 14. The red regions indicate our selection boundaries for Mira variables. The magenta and orange contours show the magnitudes and colours of known Miras and semi regular variables (SRVs) in the LMC. We also show where QSOs and AGN appear in the these diagrams with the blue contours. In the bottom panels we only include stars that show evidence of variability from the \textit{Gaia} data (see Section \ref{sec:var}).}
  \end{center}
 \end{figure*}

\section{Mira Variable Selection with 2MASS+WISE+\textit{Gaia}}
\label{sec:sel}
In this Section, we outline our selection of Mira variables in the vicinity of the LMC. In order to identify these variable stars, we make use of three all-sky surveys: the first data release (DR1) from the \textit{Gaia} mission \citep{gaia16}, the Two Micron All Sky Survey (2MASS), and the Wide-field Infrared Survey Explorer (WISE) ALLWISE catalog.

We select sources present in all three surveys that are unlikely contaminated by a nearby object. Potential artifacts are removed by imposing the 2MASS quality flags \texttt{cc\_flg=000} and \texttt{gal\_contam=0}, and the WISE flag \texttt{ext\_flg $\le 1$}. The 2MASS sources are matched to the \textit{Gaia} catalog by searching the nearest neighbours within 10". The ALLWISE catalog is then matched with the 2MASS entries by searching for neighbours within a 3" aperture.

In the following, we only consider stars with Galactic latitude $|b| > 10^\circ$ that do not suffer severely from extinction ($E(B-V) < 0.25$)\footnote{Note that we do not impose this extinction limit for stars within the inner few degrees of the LMC.}. We correct apparent magnitudes for the effects of dust extinction using the \cite{schlegel98} dust maps. Note that we use the dust reddening correction, $A_{G}=2.55 \mathrm{E}(B-V)$ for the Gaia $G$-band (see \citealt{rrl}). By comparing the de-reddened SDSS $r$ band to the uncorrected $G$ band, we find that this extinction relation holds true in the colour ranges used below.

\subsection{Magnitude and Colour Selection}
\label{sec:cmd}

In the top-left panel of Fig. \ref{fig:cmd}, we show the $G, G-J$ colour magnitude diagram for stars within 12 degrees of the LMC. The $G$ magnitude is the broad band photometry from \textit{Gaia}. The Hess diagram shows the difference between stars belonging to the LMC (black shading) and the ``foreground'' (white shading). Note that here, and throughout the paper, we define the foreground as any source not associated with the LMC. For example, the contaminants could be dwarf stars in the disc, nearby halo stars or even QSOs. The sequence of giant stars in the LMC are clear at red $G-J$ colours in the magnitude range $14 < G < 16$. The solid red lines indicate our selection of giants in the $G, G-J$ plane: $1.8 < G-J < 3.6$, $14 < G < 16$.

The top-middle and top-right panels show colour-colour diagrams using the 2MASS and WISE infrared photometry. \cite{koposov15} showed that the combination of 2MASS and WISE photometry is a powerful way of selecting giant stars in the Sagittarius stream. Here, we follow a similar procedure to identify giants associated with the LMC. The $H-K, J-H$ and $W1-W2, J-K$ colour-colour diagrams allow us to identify the LMC sequence, and we show our selection regions with the solid red lines. The polygons defining these selection boundaries are given below:
\newpage
\begin{equation}
\begin{split}
  H-K &=& [1.5,  0.3,  0.05, 0.05 , 1.5] \\
  J-H& =& [1.4,  0.7,  0.7,  0.9,  2.3] \\
   & & \\
  W1-W2&=&[1.0,  -0.2,  -0.2,  0.4,  1.2]\\
   J-K & =& [4.0,  2.0,  0.9,  0.9,  2.5]
  \end{split}
\end{equation}  

The magenta and orange contours in the top three panels of Fig. \ref{fig:cmd} show the magnitudes and colours of known Miras and semi regular variables (SRV) in the LMC \citep{soszynski09}. The contours are chosen to encompass $\sim90$ \% of the objects. Note that our definition of ``Mira variable'' in this work encompasses all types of long period variables, so our selection will also include SRVs (see Section \ref{sec:var}). We also show the QSO and AGN distribution \citep{veron10} with the blue contours: there is very little QSO/AGN contamination within our magnitude and colour selection regions.

At a fixed distance, variable giant stars can exhibit a large apparent magnitude spread due to their periodic changes in brightness and wide-range of absolute magnitudes. Indeed, the known long period variables in the LMC shown in Fig. \ref{fig:cmd} cover a large G-band magnitude range. In this paper, we do not attempt to assign distances to our selected stars, but note that our magnitude range, $14 < G < 16$, likely covers a broad distance range around the LMC (see Section \ref{sec:subs}). Furthermore, although our colour and magnitude selection will include stars belonging to the SMC (see Fig. \ref{fig:lms_bms}), our magnitude limit of $G=16$ likely excludes some of the more distant stars associated with the SMC.

The bottom panels of Fig. \ref{fig:cmd} show the magnitude and colour distributions when we only include stars that show evidence of variability (see Section \ref{sec:var}). The LMC sequences become even clearer with these cuts, and we can see that our magnitude and colour selection is efficient at selecting giant stars associated with the LMC. As we show in the next section, the addition of variability information allows us to select a very pure sample; this is vital in order to identify giant stars in the outskirts of the LMC, where the foreground stellar density is dominant.

\subsection{Variability Information from \textit{Gaia}}
\label{sec:var}

The \textit{Gaia} satellite scans across objects repeatedly during its lifetime. In the first data release, sources are observed several tens to hundreds of times. We exploit these repeat observations to identify stars that show evidence for variability. We define the ``Variability Amplitude'' as:

\begin{equation}
  A = \sqrt{N_{\rm obs}} \, \sigma(F)/F
\end{equation}

Here, $N_{\rm obs}$ is the number of CCD crossings, and $F$ and $\sigma(F)$ are the flux and flux error, respectively. This parameter can help identify stars that show large differences in Flux between repeat observations, which are above the expected noise variations.

We impose a cut of $N_{\rm obs} > 70$ in order to limit the number of spurious detections of variability. \textit{Gaia's} focal plane contains an array of 9 CCDs, hence the limit of $N_{\rm obs} > 70$ excludes stars that have been observed less than $\sim8$ times. We also exclude stars with high astrometric noise excess, $\mathrm{log_{10}}$ AEN $> 0.1$. This ensures the sources are astrometrically well behaved, and further reduces the inclusion of spurious detections. Finally, we also exclude stars with high proper motion ($\mu > 25$ mas/yr)\footnote{Note that we illustrate a more aggressive proper motion cut using M-giants in Figure \ref{fig:lms_bms}.}. These proper motions were calculated by cross-matching the \textit{Gaia} DR1 sources with the 2MASS catalog. The uncertainty in these proper motion measurements are estimated using spectroscopic QSOs, and, for the magnitude range under consideration, we estimate errors of $\sim 5$ mas/yr.

\begin{figure}
  \begin{center}
    \includegraphics[width=8.5cm, height=4.25cm]{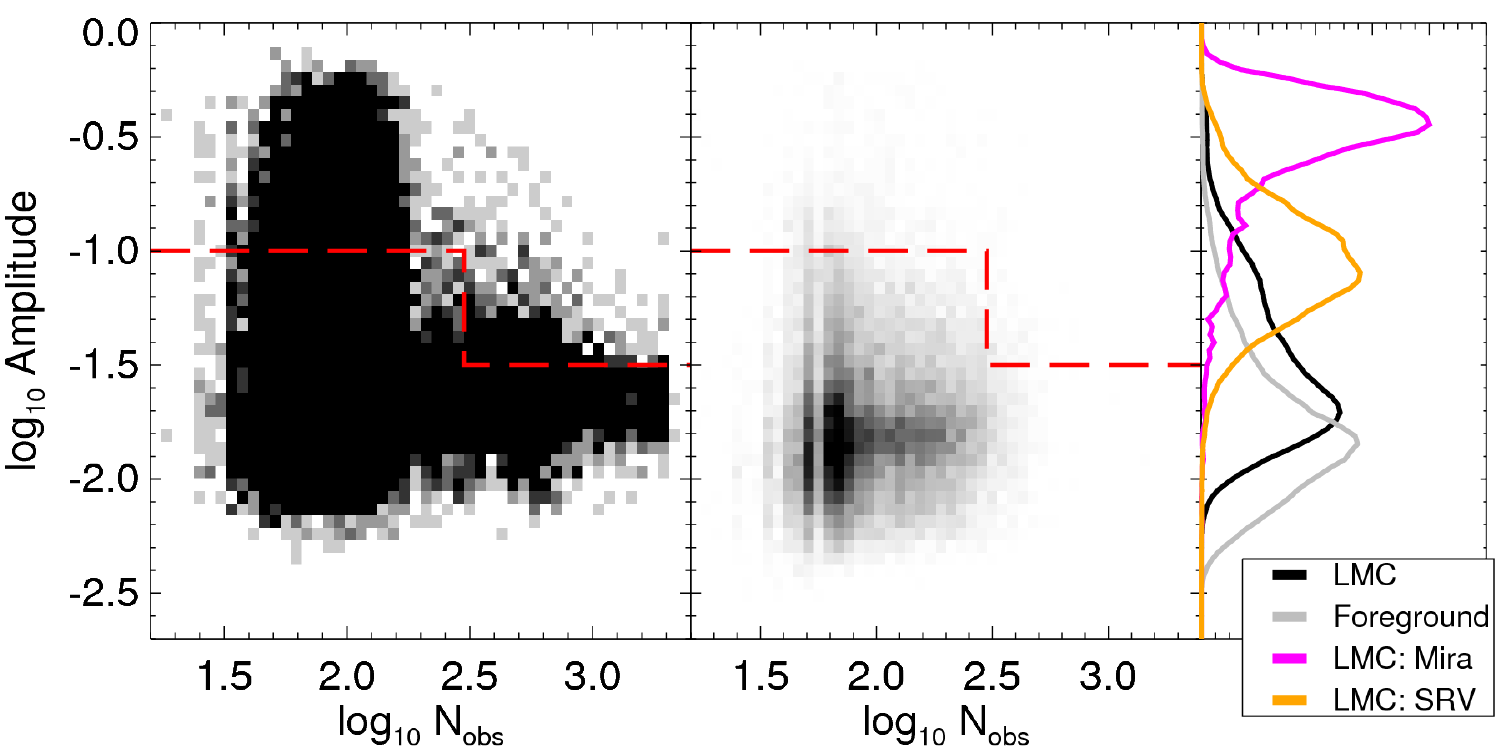}
  \end{center}
  \caption{\label{fig:amp_nobs} \small Variability amplitude ($A$) against number of observations ($N_{\rm obs}$) in the \textit{Gaia} DR1 data. The left-hand panel shows stars within 25 degrees of the LMC. The middle panel shows stars $> 25$ deg from the LMC, and $>10$ deg from the SMC. The foreground (middle panel) has been scaled to the same area as the LMC field. In both panels, only stars that are colour-selected as giants are included (see Section \ref{sec:cmd}). Note that pixels are saturated (black) at $N=5$ stars per pixel. At high amplitudes ($\mathrm{log} A > -1$) we can cleanly select variable stars in the LMC. At high $N_{\rm obs}$ the amplitudes of the long period variables decrease, and the number of contaminants decrease, so we reduce our variability threshold to $\mathrm{log} A > -1.5$. The right-hand insert shows the (normalised) distribution of amplitudes for stars in the region of the LMC (black), the foreground stars (grey), and known Miras (magenta) and SRVs (orange) in the LMC.}
\end{figure}

\begin{figure}
  \begin{center}
    \includegraphics[width=8.5cm,height=8.5cm]{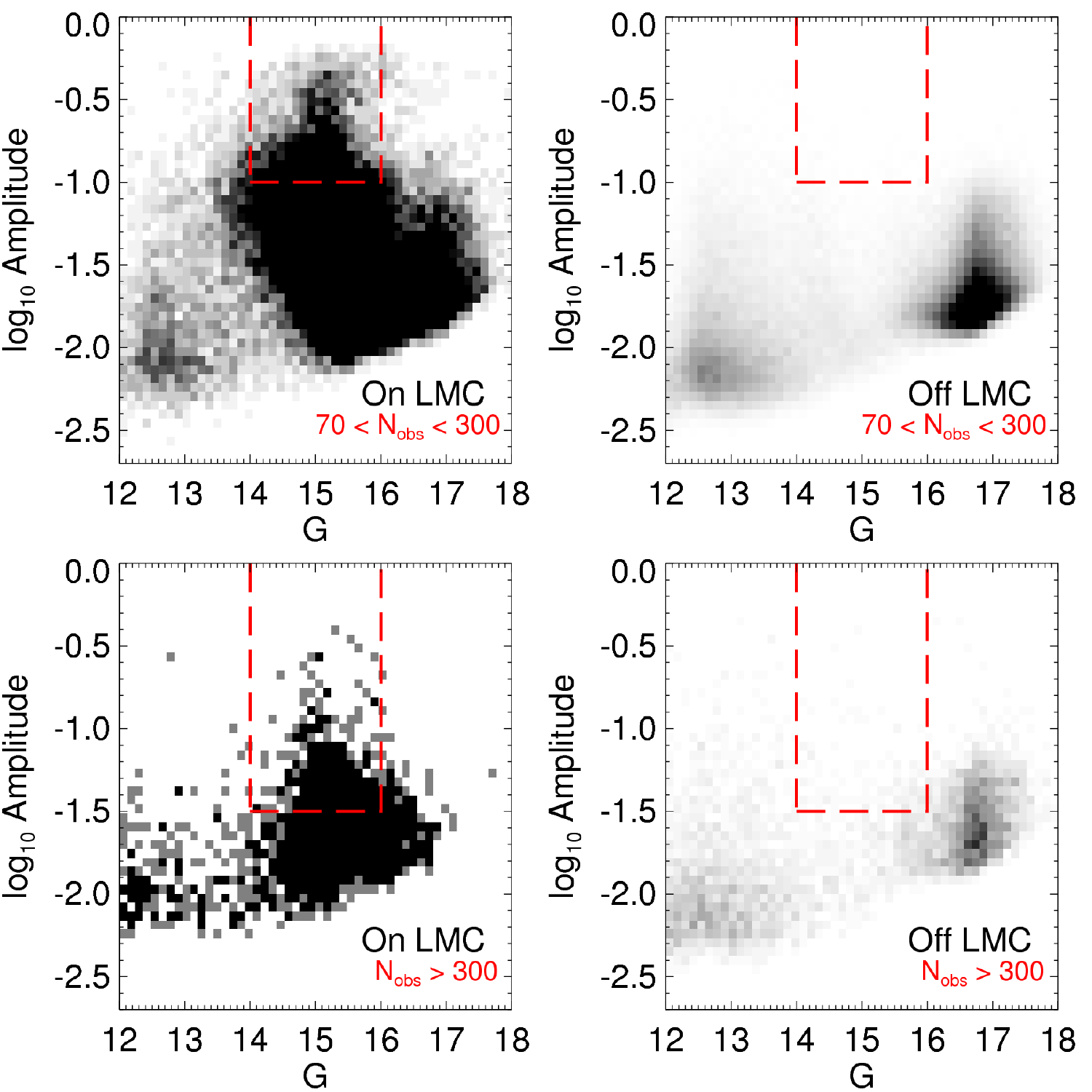}
    \caption{\label{fig:amp_gmag} \small Variability amplitude against G-band magnitude. We require a minimum threshold of $N_{\rm obs} > 70$ to determine variability, and we only include stars that are colour-selected to be giants (see Section \ref{sec:cmd}). The top panels show stars with $70 < N_{\rm obs} < 300$. The bottom panels are for $N_{\rm obs} > 300$. Pixels are saturated at $N=20 (2)$ stars per pixel in the top (bottom) panels. We show stars within 25 degrees of the LMC in the left-hand panels (``On LMC''). Objects with angular separations greater than 25 degrees from the LMC, and outside of 10 degrees of the SMC are shown in the right-hand panels (``Off LMC''). The foreground distributions have been scaled to the same area as the LMC field. Our variability thresholds and G-band magnitude range are shown with the dashed red lines: the level of contamination in this regime is very low.}
  \end{center}
 \end{figure}

\begin{figure*}
  \begin{center}
  	\begin{minipage}{0.45\linewidth}
	\centering
      \includegraphics[width=8cm, height=16cm]{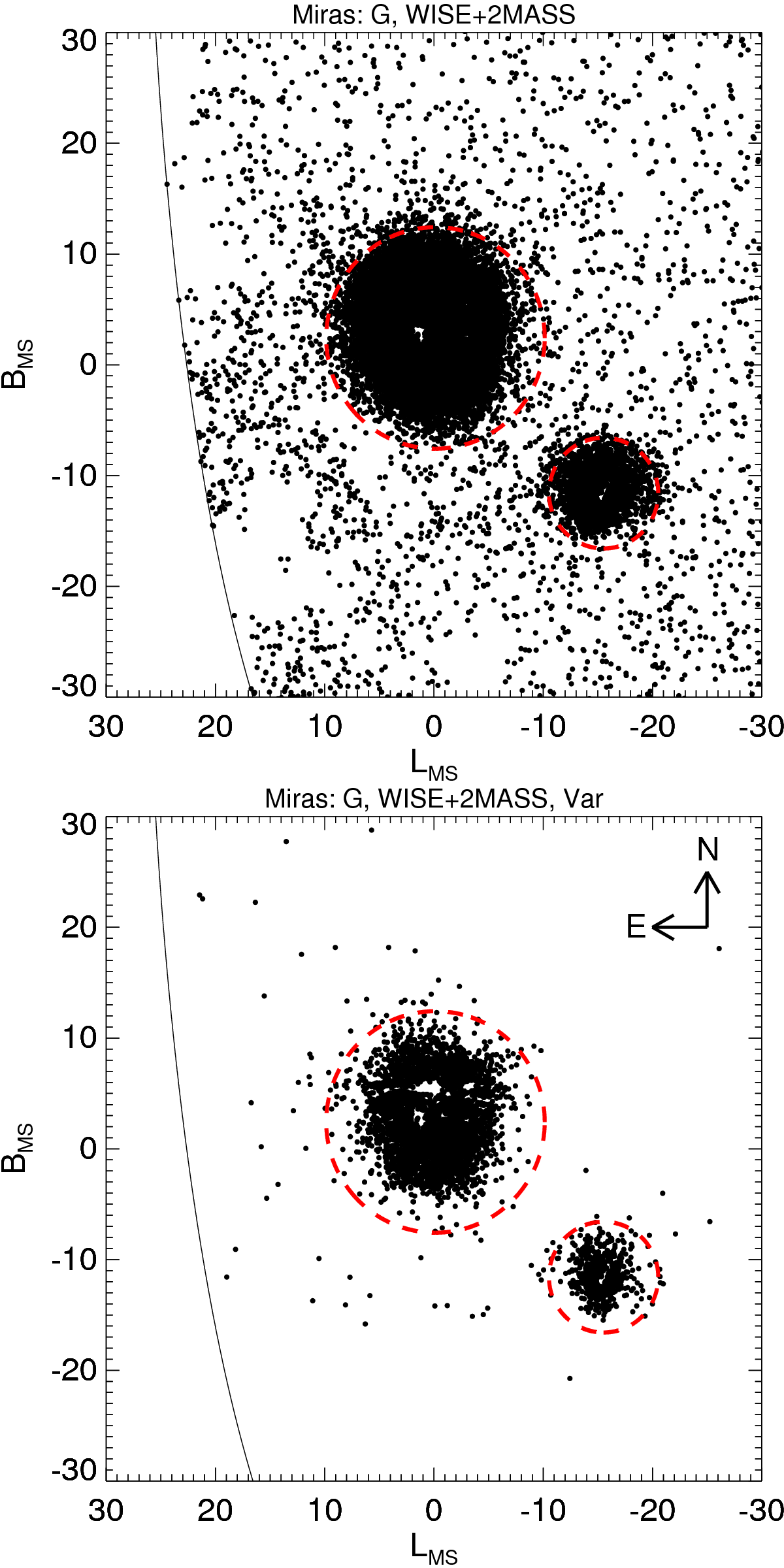}
      	\end{minipage}
	\begin{minipage}{0.45\linewidth}
	\centering
	\includegraphics[width=8cm, height=16cm]{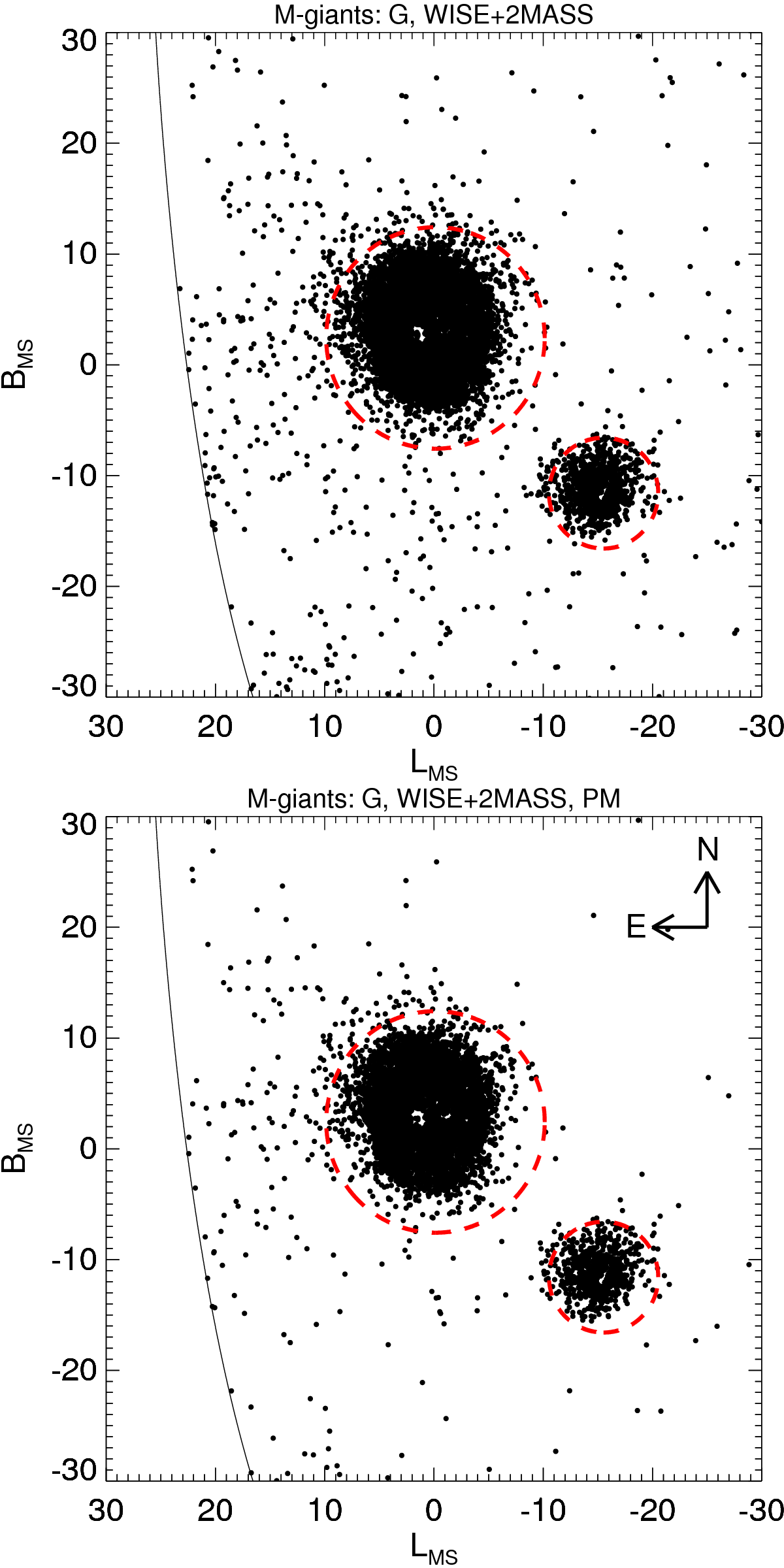}
	\end{minipage}
      \caption[]{\label{fig:lms_bms} \small Stars in the vicinity of the LMC shown in Magellanic Stream coordinates. The dashed red lines show angular separations of 10 deg and 5 deg from the LMC and SMC, respectively. The solid black line indicates a Galactic latitude of $b=-10^\circ$. In the top-left panel, stars are selected with the magnitude and colour criteria shown in Fig. \ref{fig:cmd}. The bottom-left panel includes a cut on variability amplitude and astrometric noise. The addition of the variability information from \textit{Gaia} significantly reduces the contamination and produces a clean sample of variable stars associated with the LMC. In the top-right panel, an M-giant colour selection is applied from \cite{koposov15}. The bottom-right panel includes a strict proper motion cut of $\mu < 7$ mas/yr. The M-giant selection is more complete than our variable stars, but suffers more from contamination.}
    \end{center}
\end{figure*}

In Fig. \ref{fig:amp_nobs} we show variability amplitude against $N_{\rm obs}$ for stars that pass our colour cuts.  Stars within 25 deg of the LMC are shown in the left-hand panel, and stars $> 25$ deg from the LMC and $>10$ deg from the SMC are shown in the middle panel. There is a significant excess of stars in the LMC with large amplitudes -- these are likely true variable stars. To select these Mira variables, we adopt a threshold of $\mathrm{log} A>-1$. This cut likely excludes some variable stars in the LMC, but ensures that we minimize contamination. At large $N_{\rm obs}$ the amount of contamination goes down. In addition, the amplitudes of the LMC stars seem to reduce. This is likely because, for stars observed many times, the characteristic variability timescale of the long period variables are larger than the interval over which the bulk of the observations were obtained. For example, for known Miras and SRVs in the LMC the average amplitudes are $\mathrm{log} A = -0.5$ and $\mathrm{log} A =-1.0$ for $70 <N_{\rm obs} < 300$, and the averages are $\mathrm{log} A = -0.7$ and $\mathrm{log} A = -1.2$ for $N_{\rm obs} > 300$. To account for the reduced amplitudes of ``real'' variables at high $N_{\rm obs}$, and the decreased contamination, we lower our amplitude threshold to $\mathrm{log} A>-1.5$ for $N_{\rm obs}>300$. Again, this limit likely excludes some variable giants, but ensures that we minimize contamination from non-variable stars.

The right-hand inset of Fig. \ref{fig:amp_nobs} shows the (normalised) distribution of amplitudes for stars close to the LMC (black) and in the foreground (grey). For our high amplitude cuts, the foreground level is low. We also show the amplitude distributions for known Miras and SRVs in the LMC with the magenta and orange lines, respectively. Our amplitude threshold is effective at picking out Mira stars, which generally have $\mathrm{log} A>-1$. Note that although we have defined our sample as Miras, a significant number of SRVs will also be included. Applying our colour, magnitude and variability cuts to the known long period variables from \cite{soszynski09} selects 16\% of the Miras and 15\% of the SRVs. Thus, we estimate that our sample completeness is roughly $\sim15\%$.

The efficiency of our variability amplitude criteria at selecting LMC stars is emphasized in Fig. \ref{fig:amp_gmag}. Here, we show the variability amplitude as a function of G band magnitude for stars within 25 degrees of the LMC (``On LMC'', left-hand panels), and stars away from the LMC/SMC (``Off LMC'', right-hand panels). These stars are all selected in the colour ranges outlined in Section \ref{sec:cmd}. We show the relations for $70 <N_{\rm obs} < 300$ and $N_{\rm obs} > 300$ in the bottom and top panels, respectively. The sequence of variable giants in the LMC is prominent in the magnitude range $14 < G < 16$. In the same magnitude range, the number of stars in the fields away from the LMC with high variability amplitude is very low. 

\begin{figure*}
	\begin{center}
      \includegraphics[width=15cm,height=5cm]{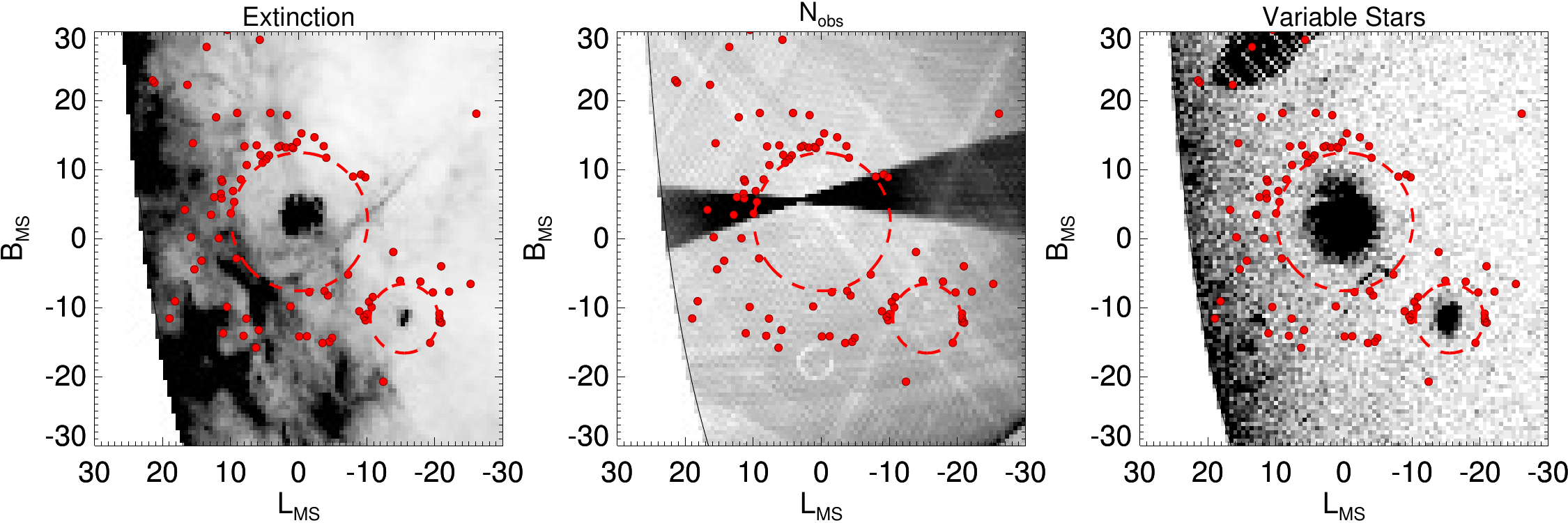}
    \caption{\label{fig:health} \small We show how extinction (left), number of observations ($N_{\rm obs}$) in \textit{Gaia} DR1 (middle) and variable artifacts (right) affect our selection of Mira variables. The range of pixel values (white to black) in each panel are: $\mathrm{E(B-V)}=[0,0.25]$, $N_{\rm obs}=[70,300]$, $N=[0,15]$. The solid black line indicates a Galactic latitude of $b=-10^\circ$. Stars within 10 angular degrees of the LMC and 5 degrees of the SMC have been omitted from these panels for clarity; the dashed red lines highlight these boundaries. There is no obvious correlation between the selected stars and extinction or $N_{\rm obs}$.  In the right-hand panel we show the density of stars in the magnitude range $14< G < 16$ with high amplitude, $\mathrm{log} A > -1$.  The over-dense region at $(L_{\rm MS}, B_{\rm MS)} \sim (10^\circ, 25^\circ)$ is likely affected by cross-match failures in \textit{Gaia} DR1, and the apparent variability of the sources is spurious. However, only a few of our candidate Miras overlap with this feature, as our selection algorithm is generally efficient at avoiding spurious sources. }
  \end{center}
\end{figure*}

In Fig. \ref{fig:lms_bms} we show how the addition of variability information from \textit{Gaia} DR1 allows us to select a clean sample of candidate Mira variables in the vicinity of the LMC. Galactic coordinates are converted to the Magellanic Stream coordinate system \citep{nidever08} by aligning with a great circle with pole $(l,b)=(188^\circ, -7.5^\circ)$. Note that in this coordinate system the LMC and SMC are centered at $(L_{\rm MS}, B_{\rm MS})=(-0.14^\circ,2.43^\circ)$ and $(L_{\rm MS}, B_{\rm MS})=(-15.53^\circ,11.58^\circ)$, respectively. In the top-left panel, we show stars in a $30\times30$ deg$^2$ area around the LMC, which have been selected with our colour and magnitude criteria. In the bottom-left panel we also apply our variability amplitude cuts. In this step, we also exclude stars with $N_{\rm obs} < 70$, $\mathrm{log_{10}}$ AEN $> 0.1$ and $\mu > 25$ mas/yr. This figure shows that the addition of variability information from \textit{Gaia} DR1 efficiently reveals structure surrounding the LMC. 

We estimate the level of contamination by applying the selection criteria outlined above to an area of sky away from the LMC. This area was chosen to (1) avoid known substructures in the halo like the Sagittarius stream (see Fig. \ref{fig:sky}), (2) have a similar \textit{Gaia} scanning law to the region surrounding the LMC\footnote{Note that the symmetric region surrounding ($l_{\rm LMC}+180^\circ, -b_{\rm LMC}$) has a similar scanning law to the LMC vicinity, where $(l_{\rm LMC}, b_{\rm LMC})=(-79.5^\circ, -32.9^\circ)$.}, and (3) have a similar Galactic latitude to the LMC. We use the region $45^\circ < l < 145^\circ$, $10^\circ < b < 60^\circ$, to  find 0.0052 stars per deg$^2$ in the foreground. Thus, within 25 deg from the LMC, the number of expected contaminants is $N=10$. Note that these contaminants are not necessarily stars that artificially show signs of variability, and could be Miras in the MW halo that are not associated with the LMC (see Fig. \ref{fig:sky}).

For comparison, we also show a selection catered towards M-giants in the right-hand panels of Fig. \ref{fig:lms_bms}. These stars are colour selected using the procedure devised in \cite{koposov15}. Specifically, we apply the following colour and magnitude cuts: $W1-W2 \le 0.03$, $0.9 < J-K < 1.5$, $0.7 < J-H < 1.3$, $-0.05 < J-K -1.2(J-H) < 0.05$, $14.5 G +6(J-K+1)<15.5$. The final cut was applied to select red giant branch stars in the distance range of the LMC. In the bottom-right panel we also apply a strict proper motion cut of $\mu < 7$ mas/yr. The M-giant selection is more complete than our sample of variable giants. However, it suffers more from contamination, especially at low latitudes close to the disc. It is worth noting that some of the broad features of our Mira distribution are also seen in the M-giants. For example, a dearth of stars in the North-West quadrant, and an excess of stars to the East of the LMC. We discuss the Mira distribution in more detail in Section \ref{sec:periph}.

In Fig. \ref{fig:health} we show the selected candidate Mira variables in the outskirts of the LMC against the distribution of extinction (left panel), and $N_{\rm obs}$ (middle panel). It is clear that our sample is not artificially tracing patterns of extinction or $N_{\rm obs}$.  In the right-hand panel of Fig. \ref{fig:health} we show the density of apparently ``variable'' sources. Here, we show all stars in the magnitude range $14 < G < 16$ with high amplitudes $\mathrm{log} A > -1$. Apart from the LMC and SMC, there is another over-dense region at $(L_{\rm MS}, B_{\rm MS}) \sim (10^\circ, 25^\circ)$. This is a portion of the sky strongly affected by cross-match failures in \textit{Gaia} DR1 (see \citealt{rrl} for more details). Reassuringly, only a few of the potential Mira stars overlap with this feature, which confirms that our selection algorithm is generally efficient at weeding out artifacts in the data.

\begin{figure}
  \begin{center}
    \includegraphics[width=8.5cm, height=6.18cm]{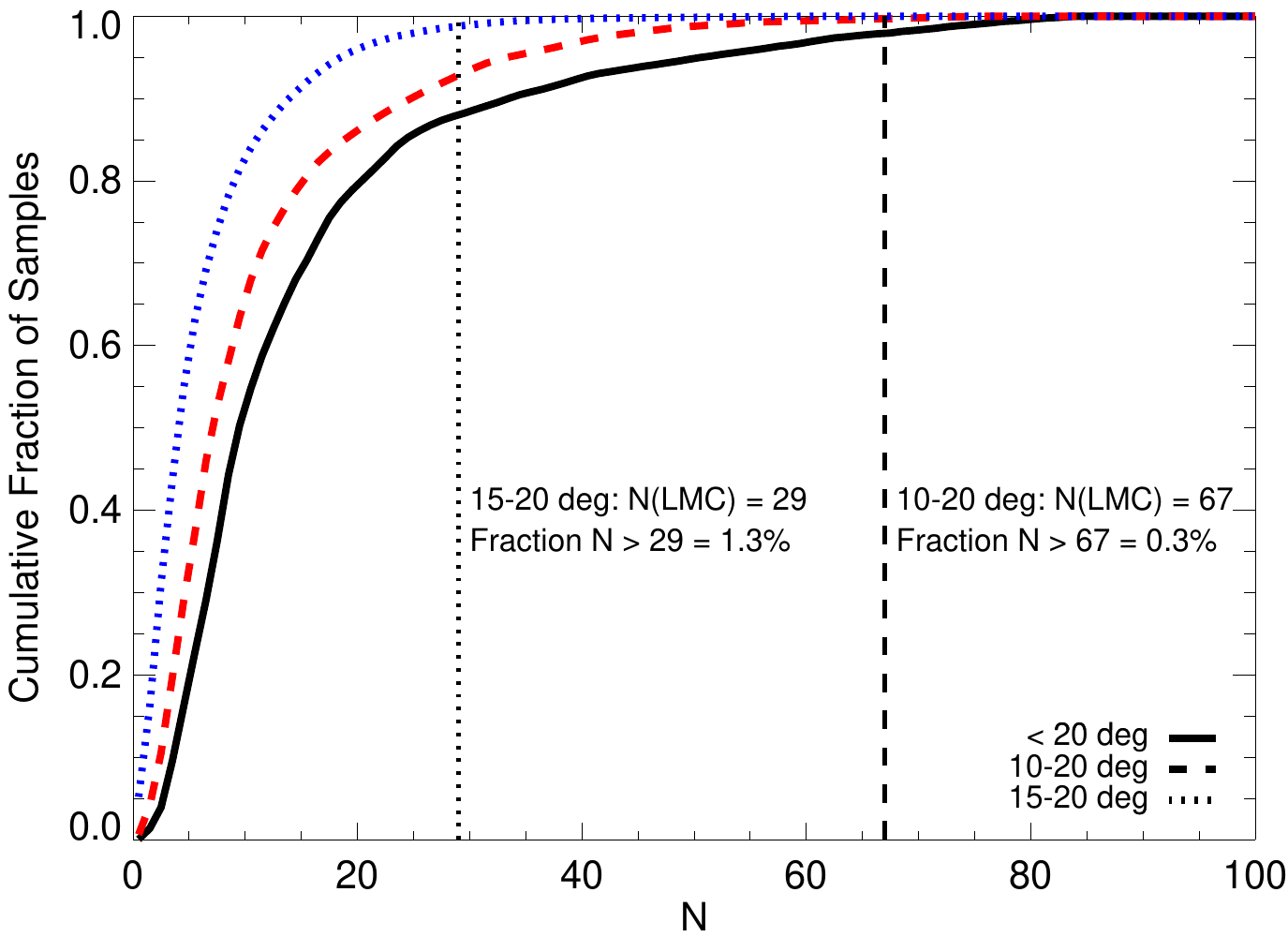}
  \end{center}
  \caption[]{\label{fig:random} \small The cumulative fraction of random pointings with $N$ candidate Mira stars within various angular separations. We show the cumulative distributions for angular separations less than 20 deg (solid black line), between 10-20 deg (red dashed line) and between 15-20 deg (blue dotted line), respectively. The pointings are randomly picked on the sky (with $|b| > 10$ deg) and the quadrant including the LMC/SMC ($l < 0^\circ, b < -10^\circ$) is excluded from the analyis. The vertical lines show the number of stars surrounding the LMC: $N=67$ between 10-20 deg and $N=29$ between 15-20 deg (excluding stars within 5 deg of the SMC).  The fraction of random samples with N $\ge$ 67(29) stars between 10-20(15-20) deg is 0.3(1.3)\%. Most of the random pointings with large $N$ are related to the Sagittarius dwarf.}
\end{figure}

Finally, in Fig. \ref{fig:random} we quantitatively compare the excess of Mira candidates surrounding the LMC to other regions of the sky. We pick $N\sim10^6$ random pointings over the sky (with $|b| >10$ deg), and exclude regions with $l < 0^\circ$, and $b < -10^\circ$ (i.e the quadrant including the LMC/SMC). For each of these pointings, the number of Mira candidates with angular separations less than $20$ deg are computed. The cumulative fraction of pointings with $N$ stars within this angular radius is shown with the solid black line in Fig. \ref{fig:random}. We also show angular separations between 10-20 deg and 15-20 deg with the dashed red and dotted blue lines, respectively. The vertical lines show the number of stars surrounding the LMC: N=67 between 10-20 deg and N=29 between 15-20 deg (excluding stars within 5 deg of the SMC). The fraction of random pointings with N $\ge$ 67(29) stars between 10-20(15-20) deg is 0.3(1.3)\%. Thus, the excess of stars in the outskirts of the LMC are unlikely due to random contamination. Furthermore, we find that the small number of pointings that do show an excess of stars that is similar to the outskirts of the LMC are all probably related to the Sagittarius dwarf. If we exclude regions of the sky likely associated with Sagittarius (i.e. the overdensities at $(l,b) \sim (5^\circ,-20^\circ)$ and $(l,b)\sim (-10^\circ,35^\circ)$ seen in Fig. \ref{fig:sky}), then we find that \textit{none} of the random pointings have N $\ge$ 67(29) stars between 10-20(15-20) deg.

In the following section, we use our clean sample of candidate Mira variables to explore structures in the outskirts of the Magellanic Clouds.

\section{Stellar Features in the Periphery of the LMC}
\label{sec:periph}

\begin{figure}
  \begin{center}
    \includegraphics[width=8.5cm, height=6.8cm]{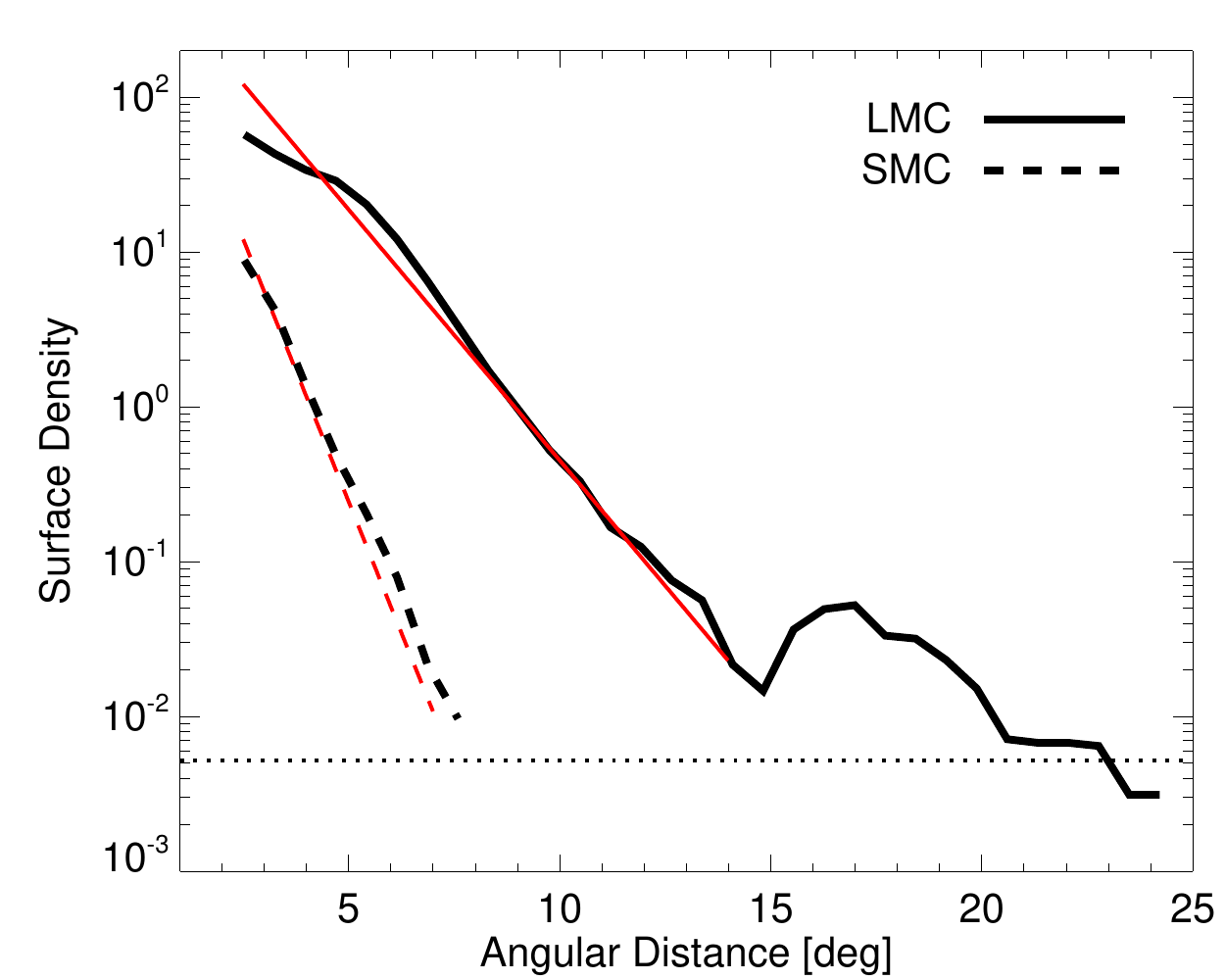}
  \end{center}
  \caption[]{\label{fig:prof} \small The surface density of candidate Mira variables as a function of angular distance from the LMC (solid black line) and SMC (dashed black line). Note that we have excluded stars within 8 deg of the SMC for the LMC profile. Approximate exponential profiles with scale lengths of 1.3 deg (LMC) and 0.6 deg (SMC) are shown with the red lines. The dotted black line indicates the approximate foreground density level (0.0052 objects per deg$^2$). The distribution of Miras in the LMC follows an extended disc distribution out to $\sim 10$ disc scale lengths,  in good agreement with the findings of \cite{saha10}. The disc appears to be truncated at $\sim 15$ deg, and the excess of stars found at larger radii are probably not associated with the ordered disc component.}
\end{figure}

\begin{figure*}
	\begin{center}
	 \includegraphics[width=15cm, height=7.5cm]{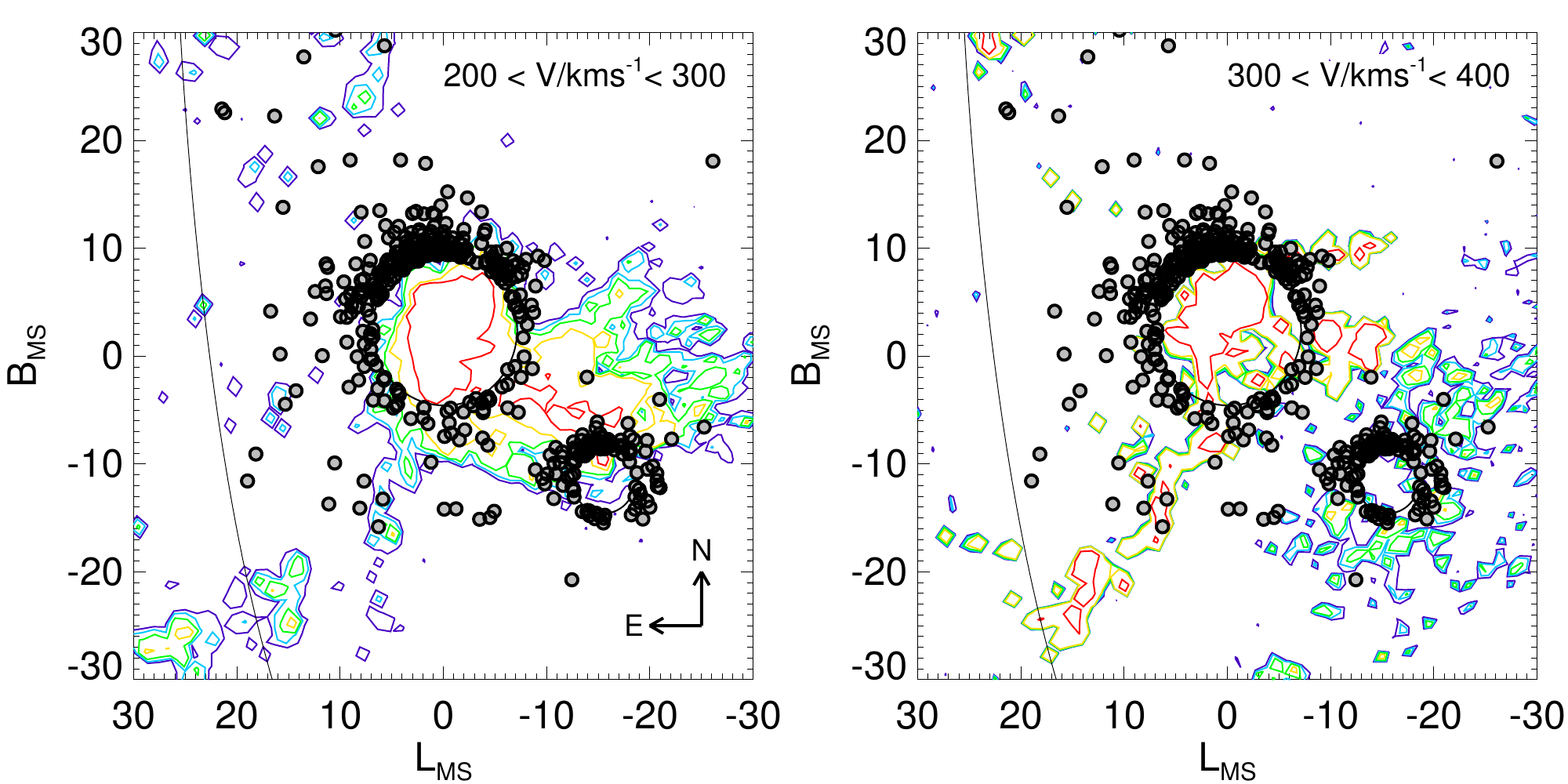}
	\caption{\label{fig:gass} \small The HI gas density in the vicinity of the LMC shown in Magellanic Stream coordinates. We apply heliocentric velocity cuts of $200 < V/\mathrm{km s}^{-1} < 300$ and $300 < V/\mathrm{km s}^{-1} < 400$ in the left- and right-hand panels, respectively. The solid black line indicates a Galactic latitude of $b=-10^\circ$. Our sample of candidate Mira variables are over-plotted with the filled gray circle symbols: stars within 7 degrees of the LMC, and 3 degrees of the SMC are omitted. The stars on the outskirts of the Magellanic system do not generally follow the gas distribution. However, there are potential stellar associations with the Leading arm in the South-East, and the ``spur'' of HI gas to the North-West of the LMC. }
      \end{center}
\end{figure*}

\begin{figure}   
\begin{center}
	 \includegraphics[width=8.5cm,height=4.42cm]{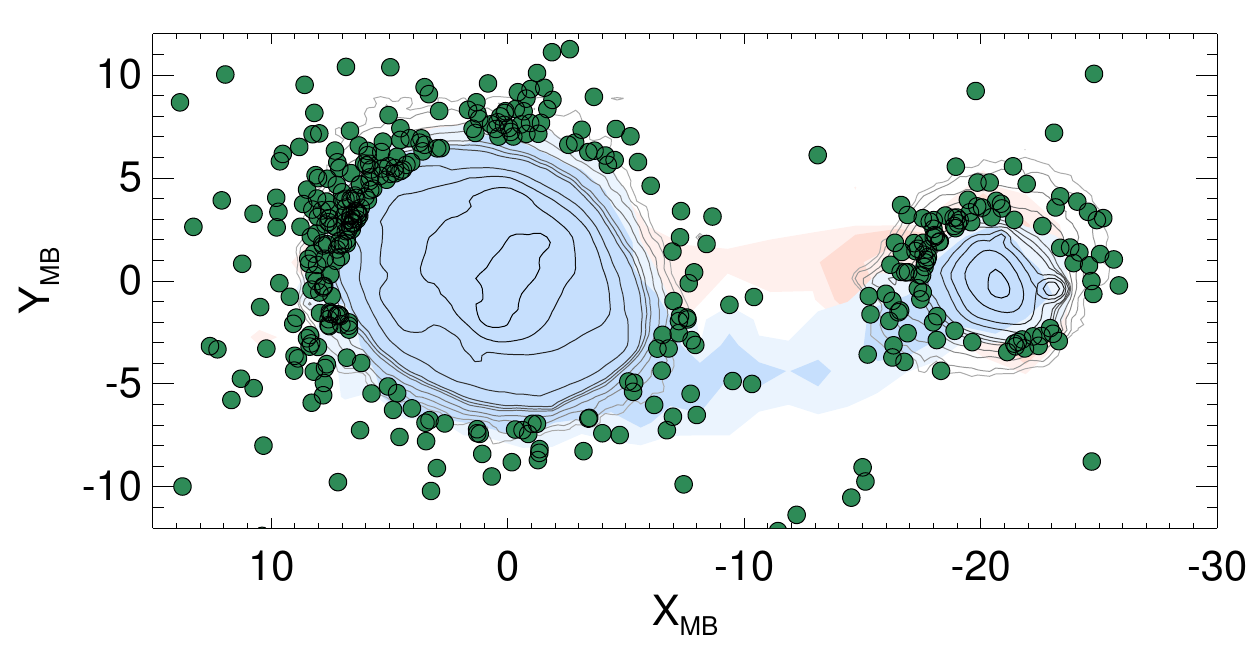}
    \caption[]{\label{fig:rrl} \small The density of RR Lyrae (RRL, blue) and Young Main Sequence (YMS, red) stars in the vicinity of the Clouds, shown in Magellanic Bridge coordinates. The stellar bridges in YMS and RRL stars are reported in \cite{rrl}. The grey contours show the counts for \textit{all} stars in \textit{Gaia} DR1. Our sample of candidate Mira variables are shown with the green filled points: stars within 7 degrees of the LMC, and 3 degrees of the SMC are omitted for clarity. The Miras do not obviously trace either stellar bridge, however the elongation of the SMC points along the RRL bridge.}
  \end{center}
\end{figure}

Our selection of candidate Mira variables allows us to explore structures associated with the LMC in regions of the sky that are dominated by the foreground stellar density. In the previous section, we found that for angular separations within $25$ deg from the centre of the LMC the level of contamination is very low. Using the contamination level of 0.0052 stars per deg$^2$, we can estimate the purity of our sample. For angular separations between 10 and 25 deg from the LMC we find 74 Mira stars (excluding stars within 5 deg of the SMC), and $N=8$ contaminants are expected. Thus we estimate that our sample purity in the outskirts of the LMC is 89\%. We explore our clean sample of peripheral LMC stars in more detail below.

In Fig. \ref{fig:prof} we show the angular profiles for the Miras surrounding the LMC and SMC. Note that stars within 8 deg of the SMC are not included in the number counts for the LMC. The distribution of Miras in the LMC looks like an extended disc with scale length $\sim 1.3$ deg, in good agreement with the findings by \cite{saha10} and \cite{balbinot15}. The disc extends to $\sim 10$ scale-lengths, and appears to be truncated at  $\sim15$ degrees from the LMC centre.  We stress that our sample is incomplete, and numbers are low at these large angular radii. However, there does appear to be a transition at $\sim 15$ degrees from the LMC centre between a disc profile and an excess of stars likely not-associated with the disc. 

\subsection{Stellar components of the Magellanic Stream?}
One of the longstanding conundrums of the Magellanic system is the existence of a stellar component associated with its vast gaseous structures. In Fig. \ref{fig:gass} we show our selected Miras in Magellanic Stream coordinates overlaid on the gas distribution surrounding the Clouds\footnote{We use the GASS Third Data Release: see \cite{mcclure09}, \cite{kalberla10} and \cite{kalberla15}.}. The contours show the density of H\textsc{i} gas with heliocentric velocity of $200 < V/\mathrm{kms}^{-1} < 300$ and $300 < V/\mathrm{kms}^{-1} < 400$ in the left- and right-hand panels, respectively. These velocity cuts were chosen to exclude the gas in the Galactic disc and emphasize the Leading Arm and Magellanic Bridge components of the gaseous material (cf. \citealt{nidever08}).

In general, the variable stars in the outskirts of the LMC do not follow the gas distribution. Notably, the stars do not appear to trace the gaseous Magellanic Bridge between the LMC and SMC. Previous work has found that the stellar component of the bridge is dominated by young ($\sim 200$ Myr) stellar populations \citep{irwin85, skowron14}, which are much younger than the intermediate age Miras.  In Fig. \ref{fig:rrl} we show the density of Young Main Sequence (YMS) and RR Lyrae (RRL) stars in Magellanic bridge coordinates. This coordinate system aligns with the great circle with pole at $(\alpha, \delta)=(39.5^{\circ}, 15.475^{\circ})$. The YMS and RRL tracers are selected using \textit{Gaia} DR1, and are described in more detail in \cite{rrl}. There is a clear bridge between the LMC and SMC in both the YMS stars \textit{and} the RRL. However, the YMS closely follows the HI gas density, but the RRL are clearly offset. 

The stellar bridge between the LMC and SMC is pronounced in RRL, but the bridge is not detected in the Mira distribution. This is likely because we are using sparse tracers, with a high level of purity, but a low level of completeness. Nonetheless, the extension of Miras at $(X_{\rm MB}, Y_{\rm MB}) \sim (-10, -5)$ near the LMC, and $(X_{\rm MB}, Y_{\rm MB}) \sim (-16, -4)$ by the SMC do point along the RRL bridge.

Finally we note that, although the candidate Miras do not generally trace the gas distribution, there are some potential associations with the Leading Arm feature of the stream in the South-East. These stars are not obviously connected to the LMC, but their distribution does bear a suggestive resemblance to the gaseous structure between the LMC and the Galactic disc. 

\begin{figure}   
\begin{center}
	 \includegraphics[width=8.5cm,height=7.3cm]{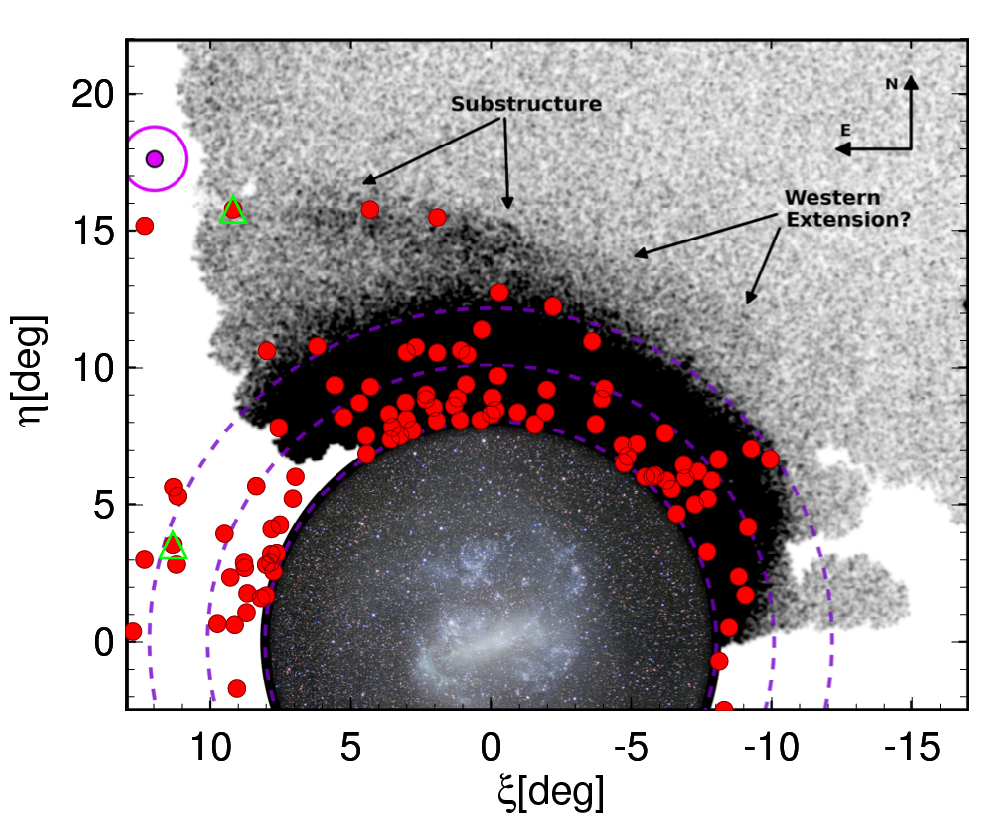}
    \caption[]{\label{fig:mackey} \small A modified version of Figure 2 from \cite{mackey16} showing the density of main sequence turn-off stars in the LMC. The coordinate system is a tangent plane (gnomonic) projection centred on the LMC. The purple dashed circles indicate angular separations of 8, 10 and 12 degrees from the centre of the LMC. Our sample of candidate Mira variables are shown with the red points. The stream identified by \cite{mackey16} is also seen in our sample, and we find that the stream likely extends further East. The purple symbol in this figure indicates the position of the Carina dwarf galaxy, and the green triangles indicate variable carbon stars identified by \cite{huxor15}.}
  \end{center}
\end{figure}

\begin{figure*}   
\begin{center}
	 \includegraphics[width=16cm,height=5.33cm]{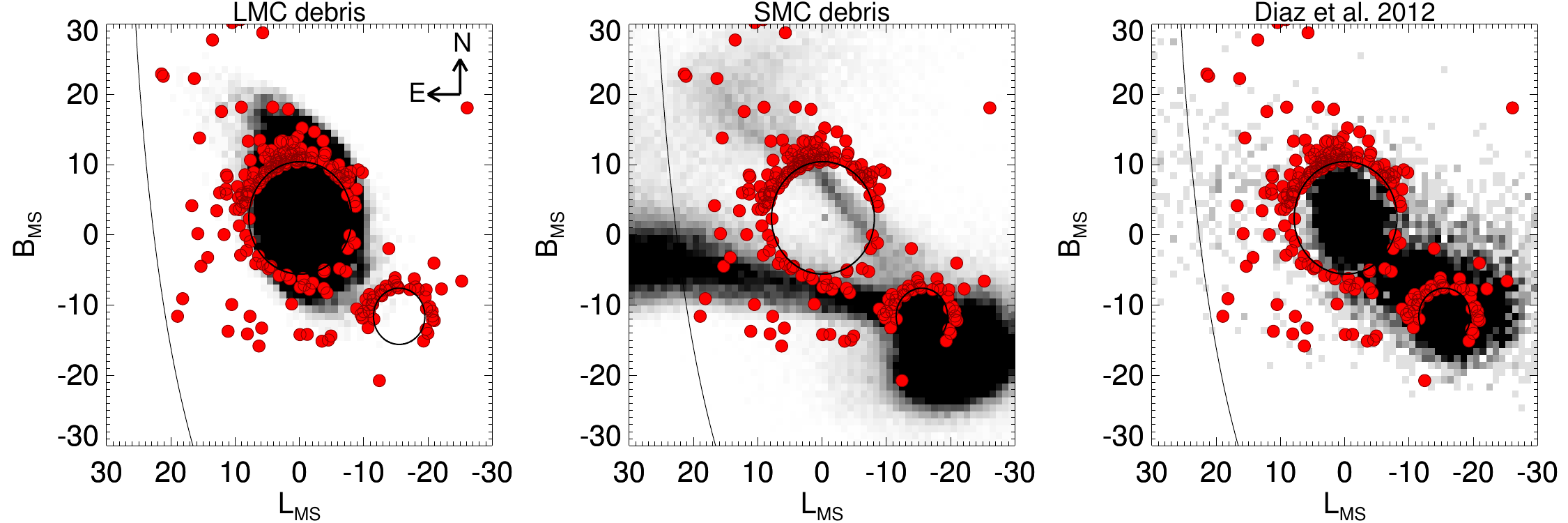}
    \caption[]{\label{fig:sims} \small The debris from simulations of SMC/LMC infall in Magellanic Stream coordinates. The solid black line indicates a Galactic latitude of $b=-10^\circ$. The left-hand panel shows the LMC debris from \cite{mackey16}.The middle panel shows the SMC debris from a large number of simulations required to match the position of the RRL stellar bridge on the sky (see \citealt{rrl}). The right-hand panel shows the debris from an SMC disruption similar to that in \cite{diaz12}.  Our sample of candidate Mira variables are shown with the red points.  Miras within 8 deg of the LMC and 4 deg of the SMC are omitted for clarity.}
  \end{center}
\end{figure*}

\subsection{Stellar features with no gaseous counterpart}

Most of the candidate Mira variables do not follow the gaseous structures of the Magellanic system. In particular, there is an excess of Miras directly East of the LMC towards the Galactic disc. One could argue that this feature is the most prone to contamination, as it is located where the foreground stellar density is very high. This could indeed be a plausible explanation for this asymmetric excess of stars. However, the density of the stars to the East of the LMC does not increase closer to the disc plane, as would be expected if the stars were associated with the Galactic disc.  Furthermore, this Eastern excess, located on the leading side of the LMC (with respect to its proper motion) is also seen in RRL stars (see Figure 16 of \citealt{rrl}), adding weight to its credibility.

An impressive stellar feature uncovered by our candidate Mira variables is an apparent stream of stars to the North of the LMC. \cite{mackey16} detected a stellar stream in this region of the sky using imaging from the Dark Energy Survey, and we recast Figure 2 from this paper in Fig. \ref{fig:mackey}. Our Mira variables are over plotted with red filled circles. The agreement with the density of main sequence turn-off from \cite{mackey16} is striking; the Miras almost perfectly delineate the stream. Furthermore, we find that the stream likely extends further East, past the limiting coverage of the DES survey.

The purple symbol in this feature indicates the position of the Carina dwarf galaxy. Previous work studying this dwarf have found evidence for stars associated with the LMC \citep{majewski00, munoz06, mcmonigal14}. As suggested by \cite{mackey16}, it is likely that these stars are part of the stream emanating from the North of the LMC. We also highlight in this figure two variable carbon stars that are present in our sample with green triangles (\citealt{huxor15}, where the IDs are HG31 and HG35). One of these stars is part of the Eastern excess described above, and the other is likely a member of the \cite{mackey16} stream.

\subsection{The origin of the outer stellar components}

\subsubsection{LMC halo stars?}
The stellar halo of the LMC likely extends over several tens of angular degrees from its centre (e.g. \citealt{belokurov16}), so the Miras in the outskirts of the LMC could be part of this halo component. However, it is not clear how an intermediate age population, that more closely resembles the stellar population of the LMC disc, could be so prevalent in the LMC halo. Furthermore, the disc of the LMC itself likely extends out to $\sim 15$ degrees (see Fig. \ref{fig:prof}, and Figure 14 of \citealt{saha10}), so the halo component may only begin to dominate over the disc at much larger radii.

\subsubsection{Stripped LMC disc stars?}
\cite{mackey16} propose that the thin stream to the North of the LMC are stripped LMC disc stars due to tidal interactions with the MW. \cite{besla16} also suggest that this stellar arc is an LMC disc feature, but the authors emphasize that such asymmetric stellar structures are more likely produced from repeated interactions between the LMC and SMC rather than MW tides. Regardless of the tidal source, the intermediate age Miras are plentiful in the LMC disc, so this is a plausible origin of the stream. However, it is unlikely that \textit{all} of the Miras we have uncovered are stripped LMC disc stars (see below and Fig. \ref{fig:sims}).

\subsubsection{Stripped SMC stars?}
It has long been argued that the gaseous material surrounding the Magellanic Clouds ought to originate from the SMC rather than the LMC (e.g. \citealt{murai80, connors06}). In all tidal models, there is a stellar counterpart to the observed gaseous component, and several models have attempted to predict both the gaseous and stellar debris from the complex three-body interaction (e.g. \citealt{besla10, besla13, diaz11, diaz12, hammer15}). For example, models by \cite{diaz12} and \cite{besla13} show that interactions between the SMC and LMC can produce stellar debris diffusely distributed behind and about the LMC disc. Furthermore, evidence for SMC populations in the LMC have already been observed \citep{olsen11}. Thus, several of our Miras could have an SMC rather than an LMC origin.\\
\\
\noindent
In Figure \ref{fig:sims} we show example simulations of stellar debris produced from the SMC/LMC infall onto the MW. We show the density of the simulated debris in Magellanic Stream coordinates, and the Miras are over plotted with the red symbols. The first simulation, shown in the left-hand panel, follows the tidal disruption of the LMC in the presence of the MW. A two-component LMC with dark matter mass $M_{200}=1.2 \times 10^{11}M_\odot$ and disk mass $M_d=4 \times 10^9M_\odot$ is simulated in the presence of a live three-component (disk, bulge and dark matter halo) MW.  This simulation was used in \cite{mackey16} to show that the observed Northern stream could be stripped LMC disc stars due to MW tides. Indeed, the material being stripped from the Northern part of the disk is coincident with the stream of Miras near the Carina dwarf. 

The other two simulations follow the disruption of the SMC in the presence of the LMC using the Lagrange cloud stripping technique of \cite{gibbons_mcls}. The middle panel shows the combined debris of 45 simulations which reproduce the stellar RRL bridge in \cite{rrl} both on the sky and in distance, and match the latest orbital constraints of the Clouds \citet{kalli13} (see \citealt{rrl} for more details). The LMC and SMC are simulated with masses of $2.5 \times 10^{11}M_\odot$ and $2 \times 10^8M_\odot$ in a three component MW potential, \texttt{MWPotential2014}, from \cite{bovy15}. The SMC material from this disruption event overlaps with the Miras to the South of the LMC and the East of the SMC. Finally, in the third panel we show a simulation that follows a similar setup to \cite{diaz12}. Here, the LMC and SMC have masses of $1 \times 10^{10}M_\odot$ and $3 \times 10^9M_\odot$, and are simulated in a three-component MW potential made up of a Miyamoto-Nagai disk, a Hernquist bulge, and an NFW halo. The setup from \cite{diaz12} was chosen to reproduce the HI features in the Magellanic Stream and Bridge. The SMC debris in this simulation wraps around the LMC and does not reach as far East as the simulation shown in the middle-panel. However, this simulation shows that it is possible for SMC material to latch onto the LMC. A less extreme ``wrap'' could explain the diffuse excess of Miras to the East of the LMC.

We only show three example simulations here, but it is clear that the SMC/LMC infall can produce both LMC and SMC debris in the vicinity of our Mira sample. It is likely that our Mira stars are a mix of stripped LMC disc stars (e.g. the thin Mackey stream) and stripped SMC stars (e.g. the more diffuse Eastern excess). Spectroscopic followup of these features will be vital in order to robustly evaluate their origin, and confront the various models of the Magellanic system. 

\section{Tracing Massive Substructures with Mira Variables}
\label{sec:subs}

\begin{figure*}
  \begin{center}
  	\begin{minipage}{\linewidth}
		\centering
   		 \includegraphics[width=15cm, height=7.5cm]{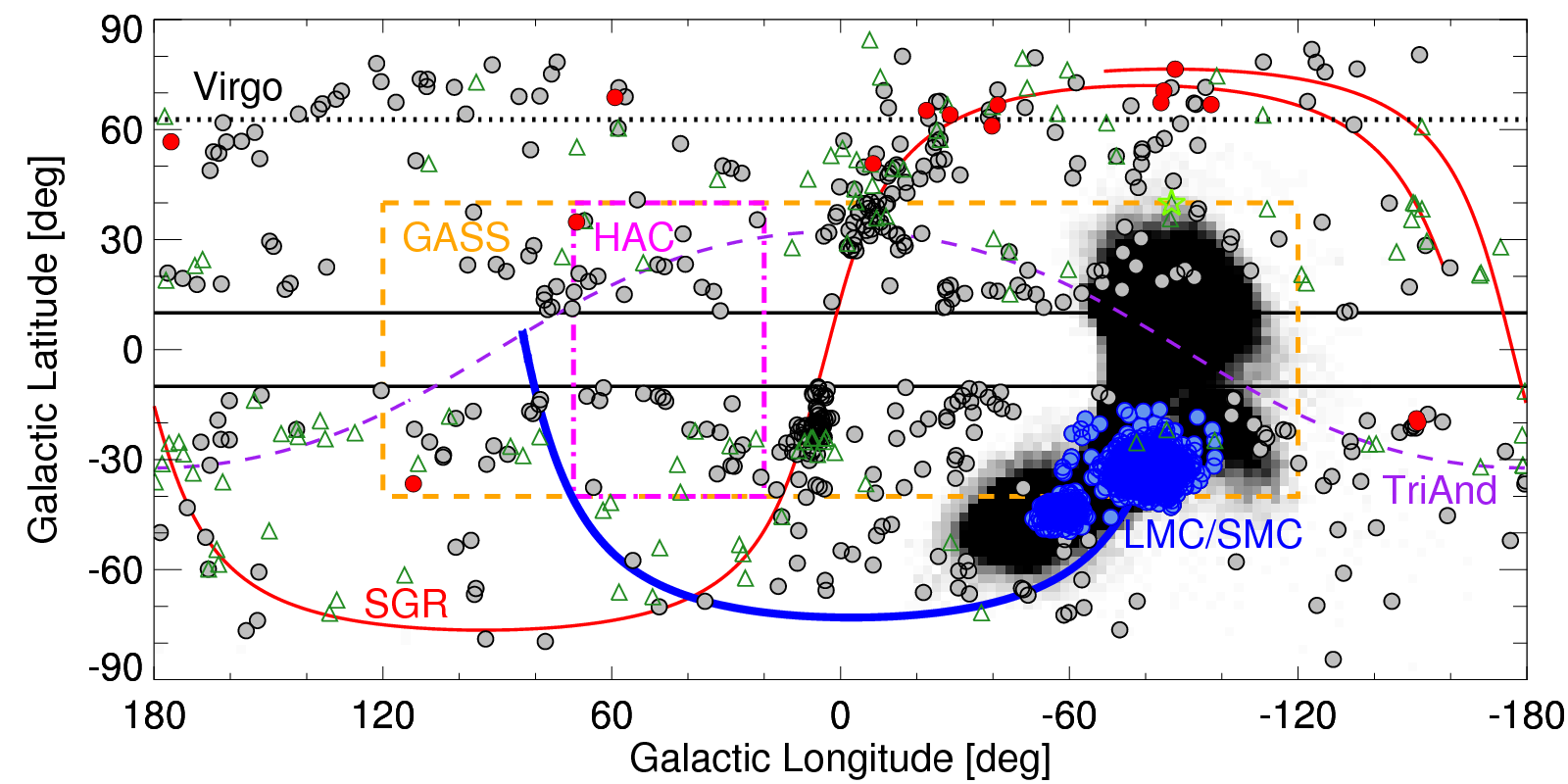}
	\end{minipage}
	\begin{minipage}{\linewidth}
		\centering
		\includegraphics[width=15cm,height=7.5cm]{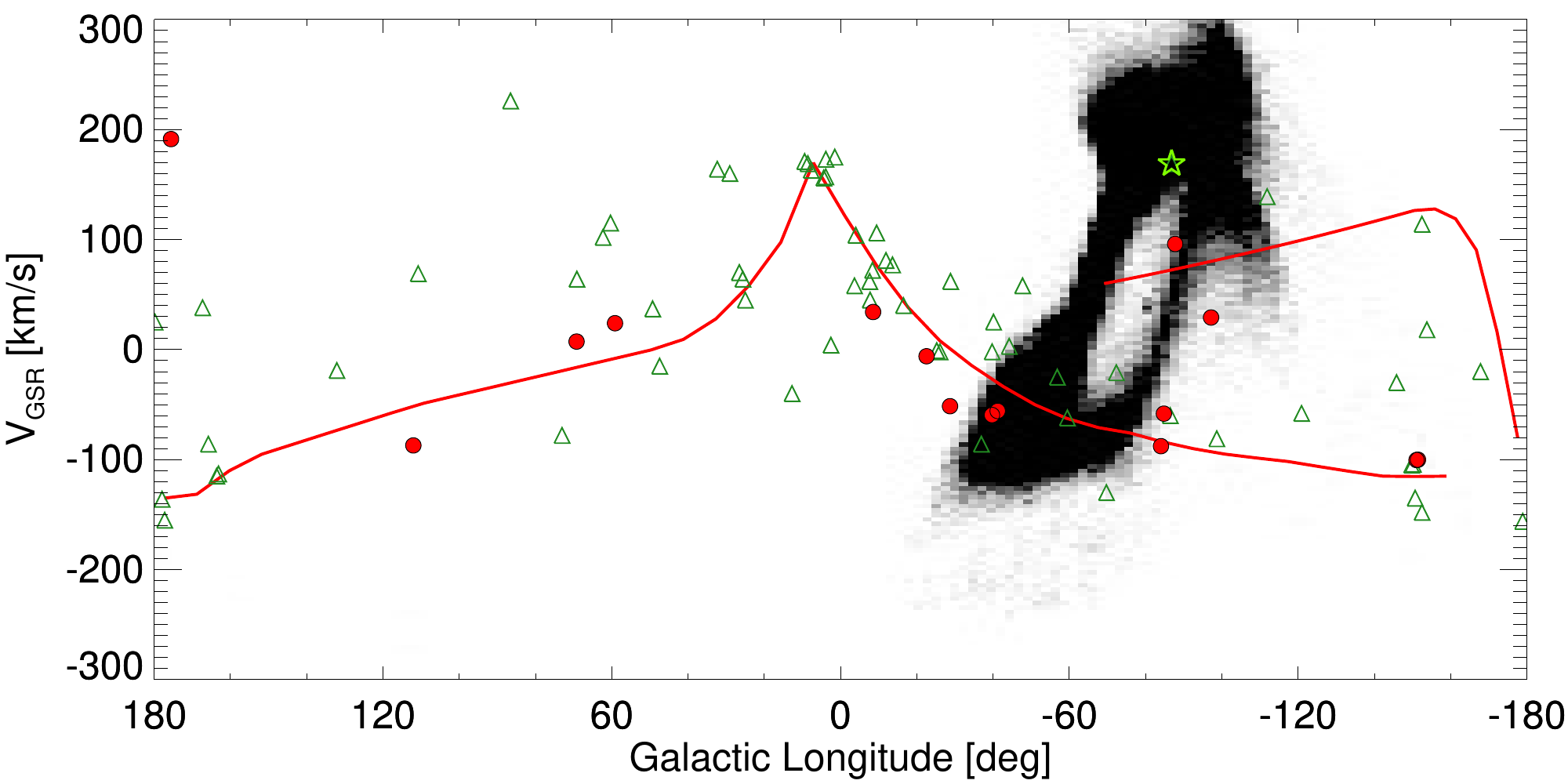}
	\end{minipage}
    \caption[]{\label{fig:sky}\small \textit{Top panel:} The distribution of candidate Mira variables in Galactic coordinates. Stars within 20 deg of the LMC and 7 deg of the SMC are highlighted in blue, and other Miras are shown with grey filled circles. Stars with spectra from APOGEE, LAMOST and SDSS are shown with red filled circles. The solid blue line shows a representative past orbit of the LMC from \cite{jethwa16}. The grey scale density shows the distribution of SMC debris from a simulation of SMC/LMC infall. Only debris in the distance range, $20 < D/\mathrm{kpc}< 70$ is shown. The red lines show the tracks of the Sagittarius leading and trailing arms \citep{belokurov14}. The approximate sky coverage of the Virgo overdensity, Hercules-Aquila Cloud (HAC) and Galactic Anti-centre Stellar Structure (GASS) are shown with the dotted black line,  dot-dashed pink line, and dashed orange line, respectively.  The dashed purple line shows a great circle through the PAndAS stream, which is likely associated with the TriAnd overdensity \citep{deason14}.  The open green triangles show the variable Carbon star sample from \cite{huxor15}. The green star symbol (located at $\left(l,b\right)=\left(-87^\circ,40^\circ\right)$) indicates a carbon star that is likely associated with the SMC debris above the Galactic plane. Most of the Miras we identify are associated with known massive substructures in the Galaxy. \textit{Bottom panel:} Galactocentric velocity as a function of Galactic longitude. The solid red lines show the tracks for the Sagittarius leading and trailing arms. The grey scale density shows the estimate velocity distribution of SMC debris from a simulation of LMC/SMC infall.}
  \end{center}
\end{figure*}

The main focus of this work is to isolate a clean sample of Mira variables in the vicinity of the Magellanic Clouds. However, given the broad-range of magnitudes probed ($14 < G < 16$), and the intrinsic magnitude spread of variable giants, our selection criteria can also be used to identify candidate Miras in the halo that are not associated with the Clouds.

In the top panel of Fig. \ref{fig:sky} we show the distribution of candidate Mira variable stars over the whole sky in Galactic coordinates. We indicate stars within 20 deg of the LMC and 7 deg of the SMC with the blue filled circles, and other Miras are shown in grey. The red lines indicate the approximate tracks of the Sagittarius (Sgr) leading and trailing arms \citep{belokurov14}. A significant fraction of the Miras away from the Magellanic Clouds are likely associated with the Sgr stream. Several authors have exploited giant stars to study the stream (e.g. \citealt{ibata02, majewski03, koposov15}), so it is not surprising that the train-wreck of the Sgr dwarf is clearly visible with the variable Mira stars.

The dashed purple line in the top panel of Fig. \ref{fig:sky} indicates a great circle through the PAndAS stream \citep{martin14}, which is likely associated with the TriAnd overdensity \citep{deason14}. We also indicate the approximate sky coverage of the Virgo overdensity ($b > +62^\circ$), Hercules-Aquila Cloud (HAC) and Galactic Anti-centre Stellar Structure (GASS) with the dotted black, dot-dashed pink, and dashed orange lines.  Many of the Miras we identify are likely associated with these known massive substructures. Our sample also traces similar structures to the variable carbon star sample compiled by \cite{huxor15} (shown in Fig. \ref{fig:sky} with the open green triangles).

Away from the known massive overdensities the number of candidate Mira variables is relatively low. This is, in part, because we do not have a complete sample, but it's also likely that the Miras are particularly effective for exploring massive, metal-rich structures. We defer this task to future work, but emphasize that our methods could easily be exploited to study other substructures in the halo.

We find that $N=15$ of our candidate Mira stars have spectroscopic measurements in the APOGEE, LAMOST and SDSS surveys. The Mira stars with spectra are highlighted in red in Fig. \ref{fig:sky}. All of the Miras with surface gravity measurements ($N=8$) are consistent with being giants (log$(g_s) < 1.15$), which confirms the purity of our sample. Most of the Mira variables with spectra are associated with the Sgr stream, and we confirm that those lying close to the stream in $(l,b)$ also have velocities consistent with the Sgr leading and trailing arms (see bottom panel of Fig. \ref{fig:sky}). This confirmation of association with Sgr allows us to estimate the approximate distance range of our sample. The Sgr core lies at $D=25$ kpc, and the leading and trailing components at $(l, b) \sim (-80,70)$ deg are located at 40 kpc and 60 kpc, respectively (see Figure 7 of \citealt{deason12} and \citealt{belokurov14}). 

As a further check, we find that $N=16$ of our sample overlaps with the \cite{huxor15} variable carbon star sample. Ten of these stars are related to the Sgr dwarf. The other stars are associated with the LMC (see Fig. \ref{fig:mackey}), the HAC, and TriAnd. Only one has no obvious association with a known substructure. The distances to these carbon stars were estimated by \cite{huxor15} using Period-Luminosity relations, and the $N=16$ in our sample span a distance range, $18 < D/\mathrm{kpc} < 63$. Thus, our sample spans the heliocentric distance range $18 < D/\mathrm{kpc} < 63$, and likely extends over a broader range of distances.\\
\\
\noindent
In Fig. \ref{fig:sky}, we reproduce the simulated SMC debris shown in the middle panel of Fig. \ref{fig:sims} with the grey scale density.  This simulation is just one realization, but it illustrates that the debris from the SMC/LMC infall can be distributed over a wide area on the sky. Several of the Mira variables we have identified in parts of the sky away from the Magellanic Clouds (e.g. above the Galactic plane at $l \sim -90^\circ$) could be associated with this disruption event. The bottom panel of Fig. \ref{fig:sky} shows that velocity measurements of stars in the region of the debris will help identify potential associations with the Clouds. Curiously, a carbon star in the \cite{huxor15} sample (with ID HG52) appears coincident with the SMC debris in both position and velocity: $(l, b, d, V_{\rm GSR}) = (-87^\circ, 40^\circ, 40 \, \mathrm{kpc}, 169 \, \mathrm{kms}^{-1})$. This star is neighbouring an excess of Mira candidates at $l \sim -90^\circ$ and $b > 10^\circ$. Thus, we speculate that these stars trace SMC debris \textit{above} the Galactic plane. 

\section{Conclusions}
\label{sec:conc}

We have exploited the first data release of the \textit{Gaia} mission to uncover candidate Mira variables in the outskirts of the Magellanic Clouds. A combination of 2MASS and WISE infrared photometry is used to select giant stars in the periphery of the LMC, and we develop a novel technique using \textit{Gaia} DR1 to identify stars showing signs of variability. Our selection of Mira variables focuses on purity rather than completeness, but allows us to probe the outskirts of the LMC with little contamination. Our main conclusions are summarised as follows:

\begin{enumerate}

\item We find candidate Mira variables surrounding the LMC out to 20 deg in all directions, apart from the North-West quadrant. The distribution of Miras outside of $\sim 10$ degrees of the LMC center do not generally follow the gas distribution. However, there are some potential associations with the Leading Arm of the Magellanic stream. 

\item We find little evidence that the intermediate-age Miras are abundant in the gaseous bridge between the LMC and SMC. However, we do detect an elongation of Mira stars in the SMC that points towards the \textit{stellar} RR Lyrae bridge discovered by \cite{rrl}. 

\item The Northern LMC stream discovered by \cite{mackey16} with main sequence turn-off stars is almost perfectly delineated by our sample of Miras. Thus, we confirm that an intermediate age population is also present in the stream, which reinforces the idea that this stream consists of stripped LMC disc stars. Furthermore, we find that the stream likely extends further East toward the Galactic plane. 

\item We find a significant excess of Mira stars to the East of the LMC, towards the Galactic plane. These stars are diffusely distributed, and are likely stripped SMC stars that have been engulfed by the LMC during past interactions between the dwarfs.

\item Our selection criteria for Miras in the vicinity of the Clouds can also be used to trace other substructures in the MW halo. Most of the Miras we identify are associated with known massive substructures in the Galaxy. In particular, the Sagittarius dSph and its associated stream are pronounced.

\item  We find that an excess of Miras \textit{above} the Galactic plane at $l \sim -90^\circ$ could be associated with the SMC/LMC disruption event. Interestingly, a known carbon star close by to these Miras is coincident in both position and velocity with a simulated distribution of SMC debris. 

\end{enumerate}

The features that our candidate Mira variables have unveiled in the outskirts of the LMC will provide important constraints on the various models of the Magellanic system. These bright tracers ($14 < G < 16$) are ripe for spectroscopic follow-up, which will be vital in order to robustly evaluate their origin. Finally, our novel technique for selecting variable giants makes use of \textit{Gaia} DR1; this is the tip of the iceberg for this unparalleled astrometric mission, which will undoubtedly prove to be a game-changer for studies of the Magellanic system, and of the Galaxy in general.

\section*{Acknowledgements}
This work has made use of data from the European Space Agency (ESA)
mission {\it Gaia} (\url{http://www.cosmos.esa.int/gaia}), processed
by the {\it Gaia} Data Processing and Analysis Consortium (DPAC,
{\small
  \url{http://www.cosmos.esa.int/web/gaia/dpac/consortium}}). Funding
for the DPAC has been provided by national institutions, in particular
the institutions participating in the {\it Gaia} Multilateral
Agreement.

A.D. is supported by a Royal Society University Research Fellowship.
The research leading to these results has received funding from the
European Research Council under the European Union's Seventh Framework
Programme (FP/2007-2013) / ERC Grant Agreement n. 308024. V.B.,
D.E. and S.K. acknowledge financial support from the ERC. S.K. 
also acknowledges the support from the STFC (grant ST/N004493/1).
We thank the anonymous referee for a constructive report.

\bibliographystyle{mnras}
\bibliography{mybib} 

\bsp
\label{lastpage}
\end{document}